\documentclass[twocolumn,nofootinbib,superscriptaddress,preprintnumbers]{revtex4}
\usepackage{amsmath,amssymb,amsfonts,slashed}
\usepackage{graphicx}
\usepackage{graphics}
\usepackage{dcolumn}
\usepackage{epsfig,color}
\usepackage{bm}
\usepackage{verbatim}
\usepackage{epstopdf}
\usepackage{appendix}

\newcommand*{\affaddr}[1]{#1} % No op here. Customize it for different styles.
\newcommand*{\affmark}[1][*]{\textsuperscript{#1}}

\begin{document}
\title{Heavy-ion double-charge-exchange and its relation to neutrinoless double-beta decay}

\author{%
E. Santopinto\affmark[1,\$], H. Garc\'ia-Tecocoatzi\affmark[1], R. I. Maga\~na-Vsevolodovna\affmark[1] and J. Ferretti\affmark[2]\\
\affaddr{\affmark[1]INFN, Sezione di Genova, via Dodecaneso 33, 16146 Genova, Italy}\\
\affaddr{\affmark[2]CAS Key Laboratory of Theoretical Physics, Institute of Theoretical Physics, Chinese Academy of Sciences, Beijing 100190, China}\\
\affaddr{\affmark[\$]santopinto@ge.infn.it}\\
%\emailnew{santopinto@ge.infn.it}\\
\affaddr{for the NUMEN Collaboration}%
}

\begin{abstract} 
We introduce the formalism to describe  heavy-ion Double-Charge-Exchange (DCE) processes in the eikonal approximation.
We focus on the low-momentum-transfer limit -- corresponding to the differential cross-section at $\theta=0^\circ$ -- and,  for the first time, we  show that it is possible to factorize the DCE cross-section in terms of reaction and nuclear parts.
Whereas in the $\theta\neq0^\circ$ case the nuclear part is a convolution of the beam and target nuclear matrix elements (NMEs), for $\theta=0^\circ$  we demonstrate - for the first time- that the transition matrix elements can be written as the sum of Double-Gamow-Teller (DGT) and Double Fermi (DF) type parts, and that they  can both be further  factorized in terms of target and projectile NMEs.
By making use of the Interacting Boson Model (IBM) formalism, we also show that the DGT  and total parts of the neutrinoless double-beta decay NMEs are in linear correlation with DCE-DGT NMEs. This confirms the hypothesis of a linear correlation between them, as  introduced in [Phys.\ Rev.\ Lett.\  {\bf 120}, 142502 (2018)].
The possibility of a Two-Step Factorization (TSF) of the very forward differential DCE-cross-section and the emergence of  DGT and DF  types for the DCE nuclear matrix elements,  combined with  a linear correlation between DCE-DGT and 0$\nu\beta\beta$ NMEs, opens the possibility of placing an upper limit on neutrinoless double-beta decay NMEs in terms of the DCE experimental data  at $\theta=0^\circ$.
\end{abstract}

\maketitle

\section{Introduction}
Neutrinoless double-beta-decay (0$\nu \beta \beta$ decay) is both one of the major experimental challenges \cite{EXO14,CUORE15,KamLAND-Zen16,GERDA17} and, at the same time, the most promising means of observing lepton-number violation. It may provide proof that neutrinos are their own antiparticles, namely that they are of the Majorana type, and information on the absolute effective neutrino mass \cite{Haxton,Doi,Tomoda, Mohapatra,Vergados}, right-handed leptonic current coupling constants \cite{Tomoda,Morales}, and also an insight in the matter-antimatter asymmetry of the universe \cite{Avignone}.
 0$\nu \beta \beta$ decay can take place in nuclei via a neutrino exchange between two quarks if the electron neutrino is a Majorana particle and has a nonvanishing mass and/or right-handed couplings \cite{Haxton,Doi,Tomoda}. 
There are also other mechanisms  which may cause the decay of two neutrons into two protons and electrons \cite{Tomoda,Mohapatra,Vergados,Avignone,Hirsch}.

The mean decay lifetime of 0$\nu \beta \beta$ processes can be calculated in terms of nuclear matrix elements (NMEs), which depend both on the weak operator and nuclear structure of the parent and daughter nuclei. 
Unfortunately, different nuclear model approaches \cite{Suhonen:2010zzc,Meroni:2012qf,Senkov:2013gso,Simkovic:2013qiy,Mustonen:2013zu,Barea} disagree in their prediction of NMEs by more than a factor of two. 
Furthermore, these results may need  additional renormalization or quenching \cite{Engel}. 
The large differences in the calculated values of NMEs give rise to some kind of theoretical error, which may severely limit the possibility of extracting the desired information on the neutrino mass once a decay signal is observed. 
These discrepancies are related to intrinsic difficulties in obtaining convergent results from many-body calculations of $0 \nu \beta \beta$ decay NMEs based on different models of nuclear structure, such as, for example, the Interacting Boson Model or the Shell Model.
Because of this, any experimental information which may help to disentangle different model descriptions may be very important. 
Some examples include two-nucleon transfer reactions \cite{twonuetron,Freeman,Kay,Entwisle,Szwec}, nuclear structure studies of parent and daughter nuclei \cite{Brown}, the study of $\beta$ \cite{Cirigliano:2013xha,Fbeta} and $2 \nu \beta\beta$ decays \cite{Barea,Barabash,Horoi,Suhonen,Rodin,Caurier}, Single-Charge-Exchange (SCE) \cite{Ichimura,Freckers,Yako,Civitarese,Nowacki,Rodriguez,Bertulani, Lenske} and pion Double-Charge-Exchange (DCE) \cite{pDCE1,pDCE2,pDCE3} reactions. 

In SCE reactions, a proton is replaced by a neutron or viceversa. SCE reactions provide information on Gamow-Teller (GT) and Fermi (F) strengths at small scattering angles, which represent another test for nuclear models and $\beta$ decay GT and F matrix elements \cite{Taddeucci}. This is related to the possibility of factorizing \cite{Taddeucci,Lenske:2018jav} the cross-section in terms of a reaction part and nuclear matrix elements, which are proportional to those involved in beta-decay processes. 

Nowadays,  several experiments on heavy-ion DCE reactions are ongoing at RNCP Osaka \cite{Osaka1,Osaka2}, RIBF RIKEN \cite{RIKEN}, and LNS-INFN \cite{Agodi:2015mta,Cappuzzello,Cappuzzello:2018wek,Cavallaro:2015pza}.
The first two of them make use of high-energy heavy-ion double-charge-exchange processes in order to study multi-spin-isospin flip excitation modes, such as a high-energy Double-Gamow-Teller  giant resonance (DGT-GR) \cite{Osaka1}, that has been predicted three decades ago \cite{Vogel,Auerbach}. 
The experiment NUMEN at LNS-INFN is aiming to extract information on DCE NMEs from heavy-ion differential cross-section, with the hope  this  can be used to put constraints  on $0 \nu \beta \beta$ decay NMEs \cite{Cappuzzello:2018wek,Cappuzzello}.

Although DCE and $0 \nu \beta \beta$ decay processes are mediated by different interactions, the former by the strong and the latter by the weak one, it has been recently proposed that  the nuclear matrix elements involved in DCE reactions may resemble, at least for their geometrical structure, those involved in $0 \nu \beta \beta$ decays \cite{Cappuzzello}. As in the SCE case, the procedure of extracting these NMEs neatly necessarily requires the factorization of the reaction and nuclear parts, a procedure which has never been demonstrated in the context of heavy-ion DCE processes. 

The aim of the present letter is to provide a theoretical description of DCE processes and, most important, to investigate the possibility of a factorization, at least within some approximations. Our first achievement: 1) is the proof that it is possible to factorize the DCE cross-section in terms of reaction and nuclear parts  for $\theta=0^\circ$; 
whereas in the $\theta\neq0^\circ$ case the nuclear part is a convolution of beam and target nuclear matrix elements (NMEs), we have shown that, for $\theta=0^\circ$, the transition matrix elements can be written 2) as the sum of DGT  and DF  parts, and 3) moreover, and most important, that they can both be further factorized, thereby disentangling target and projectile NMEs.  Finally, thanks to the Two-Step Factorization (TSF) of the very forward differential DCE-cross-section and the emergence of a linear  correlation between DCE-DGT and 0$\nu\beta\beta$ NMEs, we open the possibility of placing an upper limit on neutrinoless double-beta-decay NMEs in terms of the future  DCE experimental data  at $\theta=0^\circ$.

\section{DCE processes}
\label{Double-charge-exchange processes}
In heavy-ion DCE reactions two protons (neutrons) are converted into two neutrons (protons) in the target, and two neutrons (protons) are converted into two protons (neutrons) in the projectile, while the mass number of the target, $A$, and of the projectile, $a$, both remain unchanged.

The nucleon-nucleon charge-exchange effective potential we consider,
\begin{equation}
	V_{\rm CE}({\vec q}) = V_{\rm OPE}({\vec q}) + V_{\rm ZR} \mbox{ },
\end{equation}
is the sum of a long- and medium-range one-pion-exchange (OPE) part \cite{Backman:1985} and an effective zero-range (ZR) contact interaction  \cite{Bertsch}.
The latter, due to many-body correlations, is written in coordinate space as follows
\begin{eqnarray}
	V_{\rm ZR}({\vec r})&=\left[c_{\rm T}(\vec{\tau}_{\rm 1}\cdot \vec{\tau}_{\rm 2} )+c_{\rm GT}(\vec \sigma_{1}\cdot \vec \sigma_{2} )(\vec{\tau}_{\rm 1}\cdot \vec{\tau}_{\rm 2} ) \right]\delta^3(\vec r)  \mbox{ },
	\label{eqn:Vsr}
\end{eqnarray}
where the values of $c_{\rm GT} = 217$ MeV fm$^3$ and $ c_{\rm T} = 151$ MeV fm$^3$ are taken from the literature \cite{Bertsch}.
%%%%%%%%%%%%%%%%%%%%%%%%%%%
\begin{figure}[htbp]
\centering
\begin{tabular}{c}
\includegraphics[width=5.5cm]{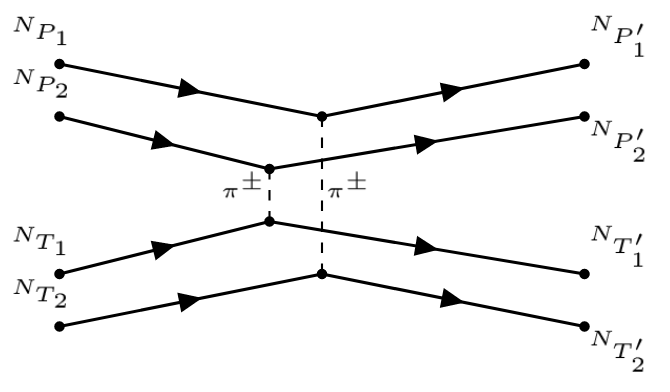}  \\ \includegraphics[width=5.5cm]{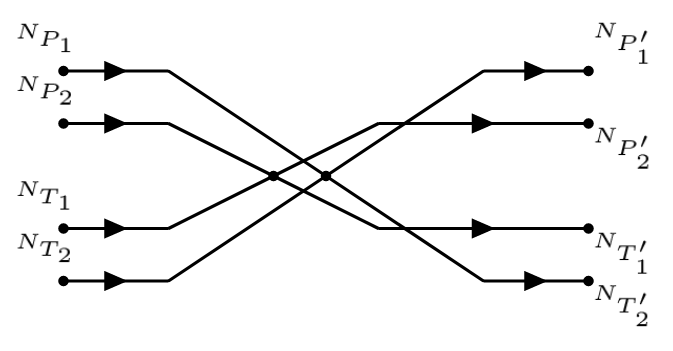}\\
\includegraphics[width=5.5cm]{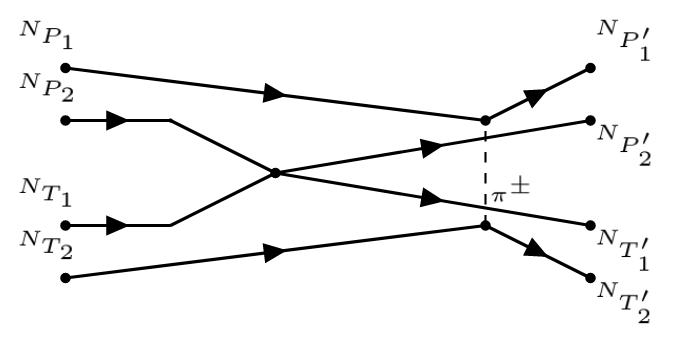}
\end{tabular}
\caption{Leading diagrams in a double-charge-exchange process. From top to bottom, they represent a double-pion-exchange interaction, a double-contact term and a mixed one-pion-exchange plus contact term.}
\label{Feyman2}
\end{figure}
%%%%%%%%%%%%%%%%%%%%%%%%%%%
The OPE and ZR interactions provide the expressions for the vertices which we need in the computation of the diagrams in Fig. \ref{Feyman2}.
These diagrams describe a DCE process in which two nucleons belonging to the target nucleus, $N_{\rm T1}$ and $N_{\rm T2}$, interact with two nucleons within the projectile, $N_{\rm P1}$ and $N_{\rm P2}$, as also depicted in Fig. \ref{fig:coordinates}.
In order to build a DCE effective potential that describes both long- and zero-range interactions, we have to combine the effects of both types of vertices. 

The DCE-effective potential that we derived in the closure approximation, where the energies of the intermediate nuclear states are replaced by an average constant value as in Refs. \cite{Haxton,Giunti}, is given by
\begin{widetext}
\begin{eqnarray}
\nonumber	V^{\rm DCE}(\vec{q}_{1},\vec{q}_{2}) & = & \frac{4}{3}\left(\frac{f_{\pi}}{m_{\pi}}\right)^4 \left( \frac{({\vec \sigma}_{\rm P1}\cdot {\vec{q}}_{1})({\vec \sigma}_{\rm T1} \cdot {\vec{q}}_{1})}{\omega_1 (\omega_1 +\bar E_{\rm P})} 
	\mbox{ } \vec{\tau}_{\rm P1}\cdot \vec{\tau}_{\rm T1} \right) \left( \frac{({\vec \sigma}_{\rm P2}\cdot {\vec{q}}_{2})({\vec \sigma}_{\rm T2} \cdot {\vec{q}_{2}})}{\omega_2 (\omega_2 +\bar E_{\rm P})(\omega_2 +\bar E_{\rm T})} \mbox{ } 
	\vec{\tau}_{\rm P2}\cdot \vec{\tau}_{\rm T2} \right)  \\ \nonumber
	&+& 2 \left[ \frac{c^2_{\rm T}}{\bar E^{\rm F}_{\rm P} +\bar E^{\rm F}_{\rm T}}  + \frac{c^2_{\rm GT}({\vec \sigma}_{\rm P1}\cdot {\vec \sigma}_{\rm T1} ) ({\vec \sigma}_{\rm P2}\cdot {\vec \sigma}_{\rm T2})}{\bar E^{\rm GT}_{\rm P}+\bar E^{\rm GT}_{\rm T}}  +\frac{c_{\rm T}c_{\rm GT} ({\vec \sigma}_{\rm P2}\cdot {\vec \sigma}_{\rm T2})}{\bar E^{\rm GT}_{\rm P}+\bar E^{\rm F}_{\rm T}} +\frac{c_{\rm T}c_{\rm GT} ({\vec \sigma}_{\rm P1}\cdot {\vec \sigma}_{\rm T1} )} {\bar E^{\rm F}_{\rm P}+\bar E^{\rm GT}_{\rm T}} \right]({\vec \tau}_{\rm P1}\cdot {\vec \tau}_{\rm T1}) ( {\vec \tau}_{\rm P2} \cdot {\vec \tau}_{\rm T2})\\
	&+& \left[ \left(\frac{f_{\pi}}{m_{\pi}}\right)^2\left( \frac{({\vec \sigma}_{\rm P1}\cdot {\vec{q}}_{1})({\vec \sigma}_{\rm T1} \cdot {\vec{q}_{1}})}{\omega_1 (\omega_1 +\bar E_{\rm P})(\omega_1 +\bar E_{\rm T})} \mbox{ } 
	\vec{\tau}_{\rm P1}\cdot \vec{\tau}_{\rm T1} \right) \big( c_{\rm T} ({\vec \tau}_{\rm P2}\cdot {\vec \tau}_{\rm T2}) + c_{\rm GT}({\vec \sigma}_{\rm P2}\cdot {\vec \sigma}_{\rm 2} )
	({\vec \tau}_{\rm P2}\cdot {\vec \tau}_{\rm T2}) \big)+ 1\leftrightarrow 2 \right]  \mbox{ },
\label{eqn:Vtpe}
\end{eqnarray}
\end{widetext}
where ${\vec \sigma}_{\rm P1,P2}$, ${\vec \sigma}_{\rm T1,T2}$, ${\vec \tau}_{\rm P1,P2}$ and ${\vec \tau}_{\rm T1,T2}$ are spin- and isospin-Pauli matrices for the projectile (P) and target (T) nucleon, $\vec q_{1,2}$ the conjugate momenta to the $\vec r_{1,2}$ coordinates of Fig. \ref{fig:coordinates}, and $\omega_i=\sqrt{{\vec{q}}_{i}^2+m_{\pi}^{2}}$, $\frac{f_{\pi}^{2}}{4\pi} = 0.08$ and $m_{\pi} = 145$ MeV are the pion coupling constant and mass, respectively. The projectile and target closure energies are given by $\bar E^{\alpha}_{\rm P}=\langle E^a_n-E^a_i\rangle_\alpha$ and $\bar E^{\alpha}_{\rm T}=\langle E^A_n-E^A_i\rangle_{\alpha}$, respectively, and the superscript, $\alpha = $ GT or F, indicates the type of energy excitation. The first line of Eq. (\ref{eqn:Vtpe}) corresponds to the double-pion-exchange contribution (first diagram in Fig. \ref{Feyman2}), the second to the double-contact term (second diagram in Fig. \ref{Feyman2}), and, finally, the third line to the mixed pion-exchange contact-term (third diagram in Fig. \ref{Feyman2}).
Further details on the derivation of the previous potential of Eq. (3) will be given in a forthcoming paper \cite{DCE-inprep}.

Here, we study the DCE transitions between $0^+$ and $0^+$ ground states for both target and projectile nuclei. In this case, the differential cross-section in the CM frame is given by
\begin{eqnarray}
\frac{d \sigma}{d \Omega} = \frac{k}{k'}\left(\frac{\mu}{4\pi^{2}\hbar^{2}} \right)^{2}|T_{\rm if}|^2 \label{cross-section}
\end{eqnarray}
where $ \mu$ is the reduced mass of the target-projectile system, $k$ and $k'$ the incoming and outgoing momentum, respectively, and $T_{\rm if}$ the T-matrix of the reaction.
We can calculate $T_{\rm if}$ by means of the Distorted Wave Born Approximation (DWBA),
%%%%%%%%%%%%%%%%%%%%%
\begin{figure}[htbp]
\centering
\includegraphics[width=8.5cm]{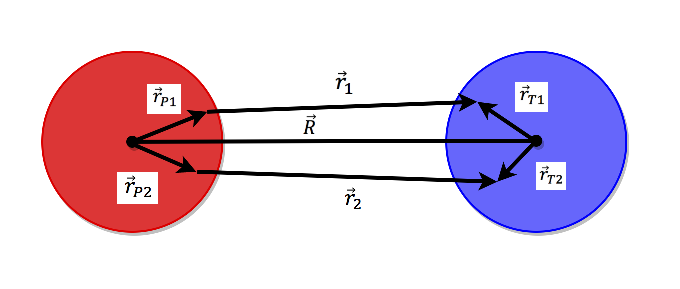} 
\caption{Coordinate system used in the calculations. $R$ is the distance between the centers of masses of the two nuclei, target (T) and projectile (P). $r_{{\rm P}1}$ and $r_{{\rm P}2}$ ($r_{{\rm T}1}$ and $r_{{\rm T}2}$) are the distances between the nucleons involved in the DCE process and the center of the projectile (target) nucleus. The coordinates ${\vec r}_{1} = {\vec R} +{\vec r}_{{\rm T}1} - {\vec r}_{{\rm P}1}$ and ${\vec r}_{2} = {\vec R} + {\vec r}_{{\rm T}2} - {\vec r}_{{\rm P}2}$ are the relative positions of the interacting nucleons.}
\label{fig:coordinates}
\end{figure}
%%%%%%%%%%%%%%%%%%%%%
\begin{subequations}
\begin{equation}
	\label{eqn:Tfi1}
	\scriptstyle
	T_{\rm if} = \left\langle \Psi^-_{{\vec k}'} \Phi_{\rm f} \right| V \left| \Psi^+_{{\vec k}} \Phi_{\rm i}\right\rangle  
	=\frac{1}{(2\pi)^{3/2}} \int  d{\vec R}
	\mbox{ }
	 {\rm e}^{i \left(\chi(b) - {\vec Q} \cdot {\vec R} \right)} M_{\rm if}({\vec m}) 
	 \mbox{ },
\end{equation} 
where we also use the eikonal approximation for the CM scattering
\begin{equation}
	 \Psi^-_{{\vec k}'}({\vec R})  \Psi^+_{{\vec k}}({\vec R}) =\frac{1}{(2\pi)^{3/2}} \text{e}^{i \left(\chi(b) - {\vec Q} \cdot {\vec R}\right)} \mbox{ },
\end{equation}
\end{subequations}
which is the product of the incoming and outgoing distorted wave functions, with momenta ${\vec k}$ and ${\vec k}'$; $\vec Q = {\vec k}' - {\vec k}$ is the momentum transferred from the beam to the target and
\begin{equation}
	\label{eqn:eikonal-phase}
	\chi(b) = - \frac{i}{\hbar v} \int_{- \infty}^{+\infty} U_{\rm opt}(z',b) dz' + i \phi_{\rm Coul} ,
\end{equation}
the eikonal phase \cite{Bertulani}, which is a function of the impact parameter, $b$, the optical potential, $U_{\rm opt}$, which describes the interaction between nuclei, and the Coulomb phase, $\phi_{\rm Coul}$.
The transition amplitude $M_{\rm if}({\bf m})$ of Eq. (\ref{eqn:Tfi1}) is given by 
\begin{equation}
 	\begin{array}{l}
 	M_{\rm if}({\bf m}) = \left\langle \Phi_{\rm f} \right| V^{\rm DCE}	\left| \Phi_{\rm i}\right\rangle \mbox{ },
	\end{array} 
	\label{AmplitudeG}
\end{equation}
where $\Phi_{\rm i,f}$ are the intrinsic wave functions of the nuclei before and after the interaction, which can be written as the product of projectile/target nucleon wave functions; the index ${\bf m} = (m_{\rm T}, m_{{\rm T}'}, m_{\rm P}, m_{{\rm P}'})$ refers to the angular momentum quantum numbers of the projectile and target nuclei  wave functions, and $V^{\rm DCE}$ is the DCE potential of Eq. (\ref{eqn:Vtpe}). In momentum space, $V^{\rm DCE}$ depends on ${\vec{q}}_{1}$ and ${\vec{q}}_{2}$, which are the momenta conjugated to the ${\vec r}_{1}$ and  ${\vec r}_{2}$ coordinates, respectively (see Fig. \ref{fig:coordinates}).

We can extract a simple and more compact form for the transition amplitude within the low-momentum-transfer limit.
Indeed, within this specific limit, $V^{\rm DCE}$ is dominated by the contact potential \cite{DCE-inprep,Ruslan-Th}, which is a zero-range interaction.
Thus, the two-nucleon-pair DCE potential of eq. (\ref{eqn:Vtpe}) can be simply written as
\begin{widetext}
\begin{equation}
	\begin{array}{rcl}
	V^{\rm DCE} & \xrightarrow[{\vec Q\rightarrow 0}]{} &2 \left[ \frac{c^2_{\rm T}}{\bar E^{\rm F}_{\rm P}+\bar E^{\rm F}_{\rm T}}  + \frac{c^2_{\rm GT}({\vec \sigma}_{\rm P1}\cdot {\vec \sigma}_{\rm T1} ) ({\vec \sigma}_{\rm P2}\cdot {\vec \sigma}_{\rm T2})}{\bar E^{\rm GT}_{\rm P}+\bar E^{\rm GT}_{\rm T}}  +\frac{c_{\rm T}c_{\rm GT} ({\vec \sigma}_{\rm P2}\cdot {\vec \sigma}_{\rm T2})}{\bar E^{\rm GT}_{\rm P}+\bar E^{\rm F}_{\rm T}} +\frac{c_{\rm T}c_{\rm GT} ({\vec \sigma}_{\rm P1}\cdot {\vec \sigma}_{\rm T1} )} {\bar E^{\rm F}_{\rm P}+\bar E^{\rm GT}_{\rm T}} \right]({\vec \tau}_{\rm P1}\cdot {\vec \tau}_{\rm T1}) ( {\vec \tau}_{\rm P2} \cdot {\vec \tau}_{\rm T2})
	\end{array} \mbox{ }.
	\label{eqn:VDCE-low}
\end{equation} 
Thus, the transition amplitude of Eq. (\ref{AmplitudeG}) within the low-momentum-transfer limit, in accordance with standard re-coupling techniques, can be re-written as
\begin{equation}
	\begin{array}{rcl}
	M_{\rm if}(\bf m) & \xrightarrow[{\vec Q  \rightarrow  0}]{} & \displaystyle 6 \sum_{J} \left\{\begin{array}{cccc} 1&1&J \\ 1&1&J\\ 0&0&0 \end{array}\right\} 
	\left\langle \phi_{\rm f}^{\rm T1} \phi_{\rm f}^{\rm P1} \phi_{\rm f}^{\rm T2} \phi_{\rm f}^{\rm P2}
	\right| \left \{ \frac{c^2_{\rm GT}(2J+1)}{\bar E^{\rm GT}_{\rm P}+\bar E^{\rm GT}_{\rm T}}  \left[ [{\vec \sigma}_{P1} \times {\vec \sigma}_{P2}  ]^J \right. \left. [{\vec \sigma}_{T1}\times {\vec \sigma}_{T2}]^{J} \right]^0 \right. \\
	& + &\left. \displaystyle \frac{c^2_{\rm T} \delta_{J,0}}{\bar E^{\rm F}_{\rm P}+\bar E^{\rm F}_{\rm T}} + \sqrt{3} c_{\rm T}c_{\rm GT}\delta_{J,0}\left(\frac{ [{\vec \sigma}_{\rm P2}\times {\vec \sigma}_{\rm T2}]^0}{\bar E^{\rm GT}_{\rm P}+\bar E^{\rm F}_{\rm T}}+ \frac{[{\vec \sigma}_{\rm P1}\times {\vec \sigma}_{\rm T1} ]^0} {\bar E^{\rm F}_{\rm P}+\bar E^{\rm GT}_{\rm T}}\right)\right\} 
	\left( \tau_{\rm T1}^+\tau_{\rm T2}^+ \tau_{\rm P1}^{-} \tau_{\rm P2}^{-} 
	\right) 
	\left| \phi^{\rm T1}_{\rm i} \phi^{\rm P1}_{\rm i} \phi^{\rm T2}_{\rm i} \phi^{\rm P2} \right\rangle  ,
	\end{array}  \mbox{ }
	\label{eqn:MDCE}	
\end{equation}   
\end{widetext} 
where the isospin-operator $\tau_{\rm T1}^+\tau_{\rm T2}^+ \tau_{\rm P1}^{-} \tau_{\rm P2}^{-}$ describes the DCE process $N_{\rm T}(A,Z)+N_{\rm p}(a,z)\rightarrow  N_{\rm T}(A,Z+2) +N_{\rm p}(a,z-2)$. To compute the transition amplitude of the $N_{\rm T}(A,Z)+N_{\rm p}(a,z)\rightarrow  N_{\rm T}(A,Z-2) +N_{\rm p}(a,z+2)$ case, the previous isospin operator must be replaced by $\tau_{\rm T1}^{-} \tau_{\rm T2}^{-}\tau_{\rm P1}^{+} \tau_{\rm P2}^{+}$.

In the present study, we focus on $0^+_{\rm i} \rightarrow 0^+_{\rm f}$ transitions of the target, for which we only have the $J=0$ contribution. In this particular case, the mixed term 
$ \sqrt{3} c_{\rm T} c_{\rm GT} ( [{\vec \sigma}_{\rm P2}\times {\vec \sigma}_{\rm T2}]^0+ [{\vec \sigma}_{\rm P1}\times {\vec \sigma}_{\rm T1} ]^0)$ in Eq. (\ref{eqn:MDCE}) vanishes because the ${\vec \sigma}$-operator is a rank-1 tensor.
Thus, Eq. (\ref{eqn:MDCE}) reduces to
\begin{eqnarray}
	\scriptstyle M_{\rm if}({\bf m}) \xrightarrow[{\vec Q  \rightarrow 0}]{}  2\left[\left(\frac{{\cal M}^{\rm DGT}_{\rm T\rightarrow T'}  {\cal M}^{\rm DGT}_{\rm P\rightarrow P'}}{\bar E^{\rm GT}_{\rm P}
	+\bar E^{\rm GT} _{\rm T}} \right)  + \left(\frac{{\cal M}^{\rm DF}_{\rm T\rightarrow T'}  \  {\cal M}^{\rm DF}_{\rm P\rightarrow P'}}{\bar E^{\rm F}_{\rm P}+\bar E^{\rm F}_{\rm T}}\right) \right]  \mbox{ }, 
	\label{amp-DGT-DF}
\end{eqnarray} 
 where  ${\cal M}^{\rm DGT}_{\rm A\rightarrow A'}$ and ${\cal M}^{\rm DF}_{\rm A\rightarrow A'}$ are  DCE-Double-Gamow-Teller (DGT) and DCE-Double-Fermi (DF) matrix elements, respectively, for a given nuclear transition of the projectile/target (A = P, T), defined as 
\begin{equation}
	\begin{array}{l}
	{\cal M}^{\rm DGT}_{\rm A\rightarrow A'} =c_{\rm GT}\left\langle \Phi^{(\rm A')}_{J'} \right| \displaystyle \sum_{n,n'} [{\vec \sigma}_{n} \times {\vec \sigma}_{n'}]^{(0)} {\vec \tau}_{n} {\vec \tau}_{n'} 
	\left| \Phi^{(\rm A)}_{J} \right\rangle  \mbox{ },
	\end{array}
	\label{eqn:DGTO}
\end{equation}
 and 
 \begin{equation}
	\begin{array}{l}
	{\cal M}^{\rm DF}_{\rm A\rightarrow A'} = c_{\rm T} \left\langle \Phi^{(\rm A')}_{J'} \right| \displaystyle \sum_{n,n'}  {\vec \tau}_{n} {\vec \tau}_{n'} 
	\left| \Phi^{(\rm A)}_{J} \right\rangle  \mbox{ },
	\end{array}
	\label{eqn:DFO}
\end{equation}
where the sum is over the nucleons ($n, n'$) involved in the process. 
Finally, the cross-section of Eq. (\ref{cross-section}) can be written in the eikonal approximation and low-momentum-transfer limit as
\begin{equation}
	\begin{array}{lll}
	\frac{d \sigma}{d \Omega} &\xrightarrow[{\vec Q   \rightarrow 0}]& \frac{k}{k'}\left(\frac{\mu}{4\pi^{2}\hbar^{2}} \right)^{2}  \displaystyle 
	\left| 2F(\theta)\left(\frac{{\cal M}^{\rm DGT}_{\rm T\rightarrow T'}  {\cal M}^{\rm DGT}_{\rm P\rightarrow P'}}{\bar E^{\rm GT}_{\rm P}+\bar E^{\rm GT}_{\rm T}}\right.\right.	 \\
	& & + \displaystyle \left.\left.\frac{{\cal M}^{\rm DF}_{\rm T\rightarrow T'}  \  {\cal M}^{\rm DF}_{\rm P\rightarrow P'}}{\bar E^{\rm F}_{\rm P}+\bar E^{\rm F}_{\rm T}}\right)\right|^2 ,
	\end{array}
\label{conexion}
\end{equation}
where the angular distribution is given by
\begin{equation}
	F(\theta) \xrightarrow[{Q_z\rightarrow0}]{} 2\pi  \int^{\infty}_{-\infty}dz \int^{\infty}_0 d b \ e^{-izQ_z} \ bJ_0 (kb \sin \theta)\ e^{i\chi(b)} \mbox{ }. \label{eqn:F-theta}
\end{equation}
The above expression is written in cylindrical coordinates, where $\vec{Q}=(\vec{Q}_t,Q_z)$ with $|\vec{Q}_t|\simeq k\ \rm sin \theta$ \cite{Bertulani}.
It should be noted that in Eq. (\ref{conexion}) the nuclear part of the differential cross-section is the sum of DGT and DF amplitudes, which are both factorized in terms of target and projectile NMEs. This will open the possibility of extracting neatly DGT and DF NMEs from DCE experimental data at $\theta=0^\circ$.

In general, the GT- and F-excitation closure energies  have different values. See \cite[Table 8]{Haxton}. In the present work, we use the values $\bar E_{\rm P}=3.38$ MeV and $\bar E_{\rm T}=5.28$ MeV for the closure energies of projectile and target nuclei, respectively. These are calculated as the average $\frac{1}{2}\left(\langle E^{A,a}_n-E^{A,a}_i\rangle_{\rm GT}+\langle E^{A,a}_n-E^{A,a}_i\rangle_{\rm F}\right)$ for both target and projectile nuclei, with mass numbers $A=40$ and $a=18$, respectively. To test the validity of our approach, we compute the $^{40}$Ca($^{18}$O, $^{18}$Ne)$^{40}$Ar DCE cross-section\footnote{In the $^{40}$Ca($^{18}$O, $^{18}$Ne)$^{40}$Ar process, projectile and target are subject to $^{18}$O $ \rightarrow ^{18}$Ne and $^{40}$Ca $\rightarrow ^{40}$Ar transitions, respectively.} at $\theta=0^\circ$ -- corresponding to the low-momentum-transfer limit -- by means of Eqs. (\ref{conexion}, \ref{eqn:F-theta}) and the $^{40}$Ca $\rightarrow^{40}$Ar nuclear matrix element reported in Table \ref{tab:results}.  
$F(\theta)$ is evaluated in the sharp-cutoff limit, where $e^{i\chi(b)}=\Theta(b-R)$, with $R=8.48$ fm corresponding to the inflection point of the modulus of $ e^{i\chi(b)}$; thus, $R$ is not a free parameter. It is estimated by means of a standard Wood-Saxon shape for the complex nuclear potential, i.e.
\begin{equation}
	\begin{array}{l}
	V_{\rm WS} = V_{\rm re} + i V_{\rm im} = - \frac{V_{\rm r}}{1 + \exp\left(\frac{r - R_{\rm r}}{a_{\rm r}}\right)} - \frac{i V_{\rm i}}{1 + \exp\left(\frac{r - R_{\rm i}}{a_{\rm i}}\right)}
	\end{array} \mbox{ }.
\end{equation}
The values of the parameters, $V_{\rm r}=-35.9$ MeV, $R_{\rm r}=8.15$ fm, $a_{\rm r}=0.43$ fm, $V_{\rm i}=-101.5$ MeV, $R_{\rm i}= 7.68$ fm, and $a_{\rm i}=0.286$ fm, are taken from Ref. \cite{WSP}.  
Our calculated cross-section is 8.9 $\mu$b/sr, which is in good agreement with the data $(8.0-10.5$ $\mu$b/sr) within the experimental error. See \cite[Fig. 2]{Cappuzzello}.
To test the importance of DF-matrix elements, we compare the differential cross-sections calculated within the $\theta =0$ limit on including the DF contribution. We get 8.9 $\mu$b/sr and 6.6 $\mu$b, respectively. We conclude that the DGT-contribution is dominant, as DF provides only a 26$\%$ correction.

\section{Results and discussion}
Our results for DCE-DGT and DCE-DF NMEs were obtained by evaluating the expectation value of spin- and isospin-operators on projectile/target nuclei wave functions; see Eqs. (\ref{eqn:DGTO}) and (\ref{eqn:DFO}).
Table \ref{tab:results} shows our results for DCE-DGT and DCE-DF projectile matrix elements in the case of $^{18}$O $\rightarrow^{18}$Ne. 
We also provide our result for the $^{40}$Ca $\rightarrow ^{40}$Ar DCE-DGT and DCE-DF target matrix elements, which we use to calculate the $^{40}$Ca($^{18}$O, $^{18}$Ne)$^{40}$Ar DCE cross-section.
The above target and projectile matrix elements are computed by means of the generalized seniority approximation, which provides a truncation scheme for the nuclear shell model \cite{Talmi:1971nhs}.
It is interesting that, in the differential cross-section at $\theta=0^\circ$, the DCE-DGT part is dominant.
Indeed, as shown in Table \ref{tab:results}, the ratio ${\mathcal M}_{\rm DCE}^{\rm P,DF}{\mathcal M}_{\rm DCE}^{\rm T,DF}/{\mathcal M}_{\rm DCE}^{\rm P,DGT}{\mathcal M}_{\rm DCE}^{\rm T,DGT}$ for $^{40}$Ca($^{18}$O, $^{18}$Ne)$^{40}$Ar is only $0.15$.
%%%%%%%%%%%%%%%%%%%
\begin{table}[htbp]
\centering
\caption{DCE-DGT and  DCE-DF matrix elements for the projectile, ${\mathcal M}_{\rm DCE}^{\rm P,DGT (DF)}$, and target, ${\mathcal M}_{\rm DCE}^{\rm T,DGT (DF)}$, calculated by means of the generalized seniority scheme. The  matrix elements are given in fm$^{-1}$.}
\begin{tabular}{cccc}
\hline
\hline
Reaction & ${\mathcal M}_{\rm DCE}^{\rm P,DGT}$ &${\mathcal M}_{\rm DCE}^{\rm P,DF}$ \\
\hline  
$^{18}$O $\rightarrow^{18}$Ne      & $0.60$ & $ 0.24$  \\
\hline
Reaction & ${\mathcal M}_{\rm DCE}^{\rm T,DGT}$ & ${\mathcal M}_{\rm DCE}^{\rm T,DF}$\\
\hline
$^{40}$Ca $\rightarrow^{40}$Ar      & $0.27$ & $0.11$   \\
\hline 
\hline
\end{tabular}
\label{tab:results}
\end{table}
%%%%%%%%%%%%%%%%%%%

Forthcoming experiments will measure the DCE cross-sections for those nuclei which are  involved in experimental $0\nu\beta\beta$ decay studies. In the following, we compute the DCE-DGT and DCE-DF target matrix elements for some of those nuclei.
Our results are reported in Table \ref{tab:results2}. These are compared with DGT, DF, and total $0 \nu \beta\beta$ matrix elements (${\mathcal M}_{0 \nu \beta\beta}^{\rm DGT}$, ${\mathcal M}_{0 \nu \beta\beta}^{\rm DF}$, and ${\mathcal M}_{0 \nu \beta\beta}^{\rm TOT}$, respectively) from Ref. \cite{Barea:2013bz}. 
In both calculations, ours and that of Ref. \cite{Barea:2013bz}, the nuclear matrix elements are computed by means of the microscopic Interacting Boson Model (IBM-2) formalism \cite{Arima:1977vie}.
%%%%%%%%%%%%%%%%%%%

\begin{table}[htbp]
\centering
\caption{Our calculated DCE-DGT (second column) and DCE-DF (third column) matrix elements for the target are compared with the $0 \nu \beta\beta$-DGT (fourth column), $0 \nu \beta\beta$-DF (fifth column) and $0 \nu \beta\beta$-total (sixth column)  matrix elements from Ref. \cite{Barea:2013bz} (with $g_{\rm A} = 1$). The matrix elements are in $fm^{-1}$.}
\begin{tabular}{cccccc}
\hline
\hline
Reaction                                          & ${\mathcal M}_{\rm DCE}^{\rm T, DGT}$ &${\mathcal M}_{\rm DCE}^{\rm T, DF}$ & ${\mathcal M}_{0 \nu \beta\beta}^{\rm T, DGT}$ & ${\mathcal M}_{0 \nu \beta\beta}^{\rm T, DF}$ &${\mathcal M}_{0 \nu \beta\beta}^{\rm TOT}$ \\ 
\hline
$^{116}$Cd $\rightarrow$ $^{116}$Sn         & $0.20$          &     $0.05$             & $0.21$   &       $-0.02$             & $0.25$ \\
$^{82}$Se  $\rightarrow$ $^{82}$Kr           &  $0.28$          &    $0.08$             & $0.31$    &       $-0.21$             & $0.50$ \\
$^{128}$Te $\rightarrow$ $^{128}$Xe         &  $0.27$          &    $0.07$             & $0.28$    &       $-0.16$             & $0.43$ \\
$^{76}$Ge $\rightarrow$ $^{76}$Se            &  $0.34$          &    $0.10$             & $0.40$    &       $-0.25$             & $0.63$ \\\hline 
\hline
\end{tabular}
\label{tab:results2}
\end{table}
%%%%%%%%%%%%%%%%%%%
Let us now study the DGT and DF matrix elements of the target.
In the target case, we  discuss the hypothesis of a linear correlation between ${\mathcal M}_{\rm DCE}^{\rm T,DGT}$ and ${\mathcal M}_{0 \nu \beta\beta}^{\rm DGT}$ or ${\mathcal M}_{0 \nu \beta\beta}^{\rm TOT}$. 
This hypothesis was introduced by N. Shimizu {\it et al.} in the context of a study aimed at finding a high-energy Double-Gamow-Teller Giant Resonance (DGTGR) \cite{Menendez}. 

In the present paper, we only deal with $0^+ \rightarrow 0^+$ DCE-DGT matrix elements of ground-state target nuclei.
Pure $0^+ \rightarrow 2^+$ DGT transitions will be the subject of a subsequent paper \cite{DCE-inprep}.
%%%%%%%%%%%%%%%%%%%
\begin{figure}[htbp]
\centering
%\left
\begin{tabular}{cc}
\includegraphics[scale=0.49]{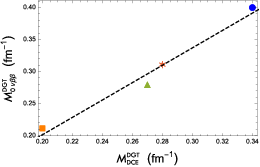}
 &\includegraphics[scale=0.49]{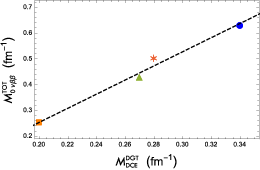}\\
 (a) & (b)
\end{tabular}
\caption{Correlation between our calculated DCE-DGT NMEs and (a) $0 \nu \beta\beta$-DGT NMEs \cite{Barea:2013bz} and (b) $0 \nu \beta\beta$-total NMEs \cite{Barea:2013bz}. The orange squares, green triangles, red stars and blue circles stand for $^{116}$Cd $\rightarrow$ $^{116}$Sn, $^{128}$Te $\rightarrow$ $^{128}$Xe, $^{82}$Se$\rightarrow$ $^{82}$Kr and $^{76}$Ge $\rightarrow$ $^{76}$Se data, respectively.}
\label{fig:correlation}
\end{figure}
%%%%%%%%%%%%%%%%%%%
Here, we conduct a simple linear regression analysis between our DCE-DGT NMEs and the $0 \nu \beta\beta$ NMEs from Ref. \cite{Barea:2013bz}, as shown in Fig. \ref{fig:correlation}.
Specifically, in Fig. \ref{fig:correlation}.a, we compare our DCE-DGT results with DGT-$0 \nu \beta\beta$ decay NMEs \cite{Barea:2013bz}. A linear correlation is seen between the two sets of data, DCE-DGT vs $0 \nu \beta\beta$-DGT. The regression line is given by 
\begin{equation}
	{\mathcal M}_{0 \nu \beta\beta}^{\rm DGT} = -0.07 + 1.36 {\mathcal M}_{\rm DCE}^{\rm T,DGT} \mbox{ }.
\end{equation}
In Fig. \ref{fig:correlation}.b, we compare DCE-DGT with $0 \nu \beta\beta$-TOT. 
The total $0 \nu \beta\beta$ NMEs can be written as \cite{Simkovic:1999re}
\begin{equation}
	{\mathcal M}_{0 \nu \beta\beta}^{\rm TOT} = {\mathcal M}_{0 \nu \beta\beta}^{\rm GT} - \left(\frac{g_{\rm V}}{g_{\rm A}}\right)^2 {\mathcal M}_{0 \nu \beta\beta}^{\rm F} 
	+ {\mathcal M}_{0 \nu \beta\beta}^{\rm T}  \mbox{ },
\end{equation}
where ${\mathcal M}_{0 \nu \beta\beta}^{\rm GT}$,  ${\mathcal M}_{0 \nu \beta\beta}^{\rm F}$ and ${\mathcal M}_{0 \nu \beta\beta}^{\rm T}$ are the Gamow-Teller, Fermi and tensor contributions, respectively; according to the hypothesis of conserved vector current (CVC), the vector coupling constant is $g_{\rm V} = 1$ \cite{Dumbrajs:1983jd}, while the value of the axial coupling constant, $g_{\rm A}$, is not defined unambiguously. 
There are three main possibilities:
\begin{equation}
	g_{\rm A} = \left\{ \begin{array}{cl} 1.269 & \mbox{ Free value \cite{Barea:2013bz,Yao:2006px}} \\ 1 & \mbox{ Quark value \cite{Rodin:2003eb,Giunti,DellOro:2014ysa}} \\ 
	1.269 \mbox{ } A^{-0.18} & \mbox{ Maximal quenching \cite{Barea:2013bz}} \end{array} \right.  .
	\label{eqn:gA}
\end{equation}
A linear correlation between the two sets of data, DCE-DGT vs $0 \nu \beta\beta$-TOT, emerges in all three cases in Eq. (\ref{eqn:gA}).
For $g_{\rm A} = 1$ (see Table \ref{tab:results2}), which is considered to be possibly the most appropriate for ${0 \nu \beta\beta}$ \cite{DellOro:2014ysa}, the regression line is given by:
\begin{equation}
	\left. {\mathcal M}_{0 \nu \beta\beta}^{\rm TOT} \right|_{g_{\rm A} = 1} = -0.29 +2.74{\mathcal M}_{\rm DCE}^{\rm T,DGT}  \mbox{ }.
\end{equation}	
If we use the free value, we have
$
	\label{eqn:reg1269}
	\left. {\mathcal M}_{0 \nu \beta\beta}^{\rm TOT} \right|_{g_{\rm A} = 1.269} =-0.17 + 2.08 {\mathcal M}_{\rm DCE}^{\rm T,DGT}  \mbox{ },
$
while in the maximal quenching case, we obtain
$
	\label{eqn:regMaxQuench}
	\left. {\mathcal M}_{0 \nu \beta\beta}^{\rm TOT} \right|_{g_{\rm A} = 1.269 \mbox{ } A^{-0.18}} =-0.78+ 5.84{\mathcal M}_{\rm DCE}^{\rm T,DGT}  \mbox{ }.
$
In conclusion, our IBM results are compatible with the hypothesis of a linear correlation between DCE-DGT and $0 \nu \beta\beta$ decay NMEs. 
This linear relation results from the short-range character of DCE and $0 \nu \beta\beta$ operators \cite{Anderson:2010aq}.
The emergence of the linear correlation is independent of the value of the axial-vector coupling constant. 
Nevertheless, different choices of $g_{\rm A}$ determine abrupt changes in the slope of the line DCE-DGT vs $0 \nu \beta\beta$-TOT.
For this reason, in order to provide more valuable information on $0 \nu \beta \beta$ decays, it will be important to place more stringent constraints on $g_{\rm A}$.
In order to do this, the effective value of the axial-vector coupling constant, $g_{\rm A}$, at the energy scale $\sim 100$ MeV relevant to $0 \nu \beta \beta$ processes \cite{Giunti,DellOro:2014ysa} should be assessed.
Several procedures have been proposed in order to extract the effective value of $g_{\rm A}$, including studies of $\beta$ and $2 \nu \beta \beta$ decays, the shape of electron spectra of forbidden $\beta$ decays \cite{Fbeta}, muon capture \cite{Suhonen-K}, and so on.

In $0^+ \rightarrow 0^+$ DCE reactions at $\theta=0^{\circ}$, a contribution is made by both DGT and DF NMEs, though the DCE-DGT contribution is the dominant one.
For this reason, one can place an upper limit on DCE-DGT NMEs, which will correspond to an upper limit  on $0\nu\beta\beta$ NMEs, thanks to the existence of the linear correlation  between them.

\section{Summary and conclusions}
We have presented the formalism for calculating  the differential heavy-ion DCE cross-sections in the eikonal approximation at very forward angles. 
We have shown explicitly, and for the first time, that, within the low-momentum-transfer limit: I) The DCE differential cross-section can be factorized into a nuclear part and a reaction factor, where the latter is computed by means of the eikonal approximation; II) The nuclear part -- which, in the case of $\theta\neq0^\circ$, is generally a convolution of the beam and  target NMEs -- can be written as the sum of DCE-DGT and DCE-DF terms, which are both further factorized in terms of target and projectile NMEs. Moreover, we have shown that the differential cross-section at $\theta=0^\circ$ is dominated by the DCE-DGT contribution; III) Analogously to the SCE case, where it was shown that the differential cross-section at $\theta = 0^\circ$ only received contributions from the contact term \cite{Bertulani}, the DCE differential cross-section is dominated by contact interactions.

In conclusion, the possibility of factorizing the very forward differential DCE-cross-section, in combination with the existence of a linear correlation between the DCE-DGT and 0$\nu\beta\beta$ NMEs, opens the possibility to place constraints on neutrinoless double-beta-decay NMEs in terms of the DCE experimental data  at $\theta=0^\circ$.

\end{document}